\documentclass[journal]{IEEEtran}
\IEEEoverridecommandlockouts
\usepackage{mathrsfs}
\usepackage{amsfonts}
\usepackage{ifpdf}
\usepackage{cite}
\usepackage{array}
\usepackage{booktabs}
\usepackage{setspace}
\usepackage{amssymb}
\usepackage{mathrsfs}
\usepackage{supertabular} 
\usepackage{multirow}
\usepackage{dcolumn}
\usepackage{booktabs}

\ifCLASSINFOpdf
\usepackage[pdftex]{graphicx}
\else \fi
\usepackage[cmex10]{amsmath}
\usepackage{array}
\usepackage{mdwmath}
\usepackage{mdwtab}
\usepackage{eqparbox}
\usepackage[tight,footnotesize]{subfigure}
\usepackage{algorithmic}
\usepackage{algorithm}
\usepackage{amsmath}
\usepackage{epsfig}
\usepackage{ae}
\usepackage{amssymb}
\usepackage{times}
\usepackage{amsmath}   

\usepackage{xcolor}

\usepackage{amsthm} 

\usepackage{supertabular} 

\usepackage{multirow}

\newlength\savewidth

\begin{document}

\title{SVM-Based Sea-Surface Small Target Detection: A False-Alarm-Rate-Controllable Approach}
\author{
         Yuzhou Li,~\IEEEmembership{Member,~IEEE}, Pengcheng Xie, Zeshen Tang,
         and~Tao Jiang,~\IEEEmembership{Senior Member,~IEEE}
\thanks{The authors are with the School of Electronic Information and Communications, Huazhong University of Science and Technology, Wuhan, 430074, P. R. China (e-mail: {yuzhouli, xiepengcheng, zeshentang, taojiang}@hust.edu.cn).}
}

\maketitle
\IEEEpeerreviewmaketitle
\begin{abstract}
In this letter, we consider the varying detection environments to address the problem of detecting small targets within sea clutter.
We first extract three simple yet practically discriminative features from the returned signals in the time and frequency domains and then fuse them into a 3-D feature space.
Based on the constructed space, we then adopt and elegantly modify the support vector machine (SVM) to design a learning-based detector that enfolds the false alarm rate (FAR).
Most importantly, our proposed detector can flexibly control the FAR by simply adjusting two introduced parameters, which facilitates to regulate detector's sensitivity to the outliers incurred by the sea spikes and to fairly evaluate the performance of different detection algorithms.
Experimental results demonstrate that our proposed detector significantly improves the detection probability over several existing classical detectors in both low  signal to clutter ratio (SCR) (up to $58\%$) and low FAR (up to $40\%$) cases.
\end{abstract}
\section{Introduction}
Accurate detection of small targets on sea surface is an important problem in remote sensing and radar signal processing applications\cite{8304386}.
However, when detecting, the radar returns from the small targets are severely obscured by the backscatter from the sea surface, which is referred to as sea clutter\cite{8304386}.
To identify the small targets from the sea clutter, a promising approach is to seek certain features from the returned signals that can depict the intrinsic differences between these two classes and then design a feature-based detector.
However, the extracted features usually become ineffective when the detection environment changes,  as the characteristics of the sea clutter are highly dependent on the sea states and radar's parameter configurations.
Therefore, extracting robust features from the returned radar signals that adapt to varying environments is crucial for target detection.

There have been extensive works to design potentially discriminative features for detecting small targets within sea clutter.
In \cite{4148430}, the authors utilized a doppler spectrum feature to describe the differences between the sea clutter and target signals, where the detector's decision was made by simply comparing the feature's value with a predefined threshold.
However, such single feature based detector only exploits limited information of the returned signals and thus its detection performance is likely to be affected by the varying detection environments.
Consider this, a potential solution for detection performance improvement is to integrate more features to construct multi-dimensional feature spaces, as by this more additional information within the returned signals can be provided.
Following this insight, Xu in \cite{5395653} extracted two temporal fractal features to devise a 2-D convexhull learning algorithm for detection.
Further, Shui $\textit{et. al}$ in \cite{6850164} introduced three features, i.e., the RAA, RPH, and RVE, to construct a 3-D feature space, under which the detection accuracy is improved in both high and low signal to clutter ratio (SCR) scenarios compared with several single feature based detectors.

Nevertheless, it should be noted that the detection performance in \cite{5395653} and \cite{6850164} is still poor in low SCR scenarios, e.g., lower than $57\%$ when SCR $=$ -2 dB.
To further promote the robustness of the detectors, the following two ideas could be considered.
Firstly, seek more discriminative features.
It was observed that some features such as the widely-adopted amplitude become ineffective in low SCR scenarios \cite{7859306}.
On the contrary, we find that some concepts in other research fields can be used to define features that are effective even in low SCR situations, e.g., the information entropy in the communication theory.
Secondly, establish more advanced detection frameworks.
Several recent works have shown that machine learning based techniques exhibit excellent potential in target detection compared with some conventional approaches \cite{islam2012artificial,7987714,7775813}.
One of their main advantages is that they can adaptively adjust the involved parameters and decision regions according to the collected radar returns, which are usually predefined in existing popular frameworks, e.g., the constant false alarm rate (CFAR) detector \cite{7737020}.
In this way, learning-based detectors may be less sensitive to the variation of the detection environments.

In view of these, this letter devotes to exploring discriminative features for feature space construction and designing a learning-based detector for accurate small target detection.
The main contributions of this work are as follows:
\begin{itemize}
\item We exploit some concepts in other research fields to define three features i.e., the temporal information entropy (TIE), the temporal Hurst exponent (THE), and the frequency peak to average ratio (FPAR), from the perspective of time and frequency domains. Particularly, the three defined features are quite simple yet practically discriminative under varying detection environments even in low SCR and false alarm rate (FAR) cases.
\item We adopt and elegantly modify the support vector machine (SVM), a classical binary classifier, to design a learning-based detector. Significantly different from the existing learning-based detectors, our proposed detector enfolds the FAR and can flexibly control it by simply tuning two introduced parameters.
    By this,  it is convenient to fairly evaluate the performance of different detection algorithms, and to flexibly regulate the sensitivity of the detector to the outliers incurred by factors such as the sea spikes to meet the requirements of different applications.
\item Experimental results show that, compared with several classical detectors, our proposed detector significantly improves the detection probability in both low SCR (up to $58\%$) and low FAR (up to $40\%$) cases.
\end{itemize}

\section{Feature Space Construction}
\label{Section:SystemModel}
In this section, we adopt the Intelligent PIxel Processing X-band (IPIX) database, a widely-used database for sea-surface small target detection, to extract features.
The IPIX database contains amount of sea clutter datasets, collected by the IPIX radar at the east coast of Canada in November 1993 \cite{websitesdata}.
In this database, each dataset is composed of 14 spatial range cells and each cell has $2^{17}$ samples with a sampling rate of 1000 Hz.
For each dataset, the cell with the target returns is labeled as the primary cell, the adjacent cells affected by the target are labeled as the secondary cells, and the remaining cells are clutter-only cells.
In addition, each dataset contains four kinds of data, referred to as the HH, VV, HV, and VH data, as the transmitter and receiver of the IPIX radar have two channels with H and V polarizations, respectively.
Throughout this letter, we will use 10 datasets in the IPIX database, namely the datasets $\#54$, $\#30$, $\#31$, $\#310$, $\#311$, $\#320$, $\#40$, $\#26$, $\#280$, and $\#17$.
For notational simplicity, we denote the samples from the primary cell and clutter-only cells as target signals and sea clutter signals, respectively, in the following.

Based on the 10 selected datasets, this section extracts three simple yet practically discriminative features from returned radar signals in the time and frequency domains, and then based on them to construct a 3-D feature space.

\subsection{Temporal Information Entropy}
We first utilize the concept of the information entropy in the communication theory to define a feature in the time domain.
Let $x = \{x_i, i = 1,2,\cdots,N\}$ be a time sequence composed by the amplitudes of the returned signals.
Divide the amplitude range covered by $x$ into $K$ ($K \in \mathbb{N^+}$) independent segments with equal length and use $N_k$ to denote the amount of the elements falling into the $k$-th segment.
Then, the probability that the amplitude of returned signals falls into the $k$-th segment, denoted by $P(N_k)$, can be calculated as
\begin{equation}\label{eq:informationEntropy_probability}
P(N_k) = \frac{N_k}{N}.
\end{equation}

Accordingly, the information entropy of such a time sequence, referred to as the temporal information entropy (TIE) in this letter, is expressed as
\begin{equation}\label{eq:informationEntropy_one}
\text{TIE}(x) = -\sum_{k=1}^KP(N_k)\log_2(P(N_k)).
\end{equation}
To avoid invalid calculation, we set $P(N_k)\log_2 P(N_k)=0$ when $P(N_k)=0$.
From the above definition, it can be interpreted that the TIE actually reflects the temporal variation or randomness of the amplitudes of returned signals.

\begin{figure}[t]
\centering \leavevmode \epsfxsize=3.5 in  \epsfbox{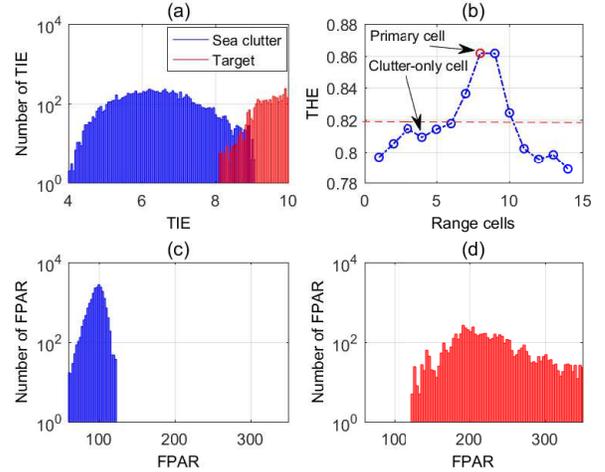}
\centering \caption{Illustrations of the extracted features. (a) The TIE of the returned signals. (b) The THE of different range cells. (c) The FPAR of the sea clutter signals. (d) The FPAR of the target signals.}
\label{Fig:Describution}
\end{figure}

\begin{figure}[t]
\centering \leavevmode \epsfxsize=3.5 in  \epsfbox{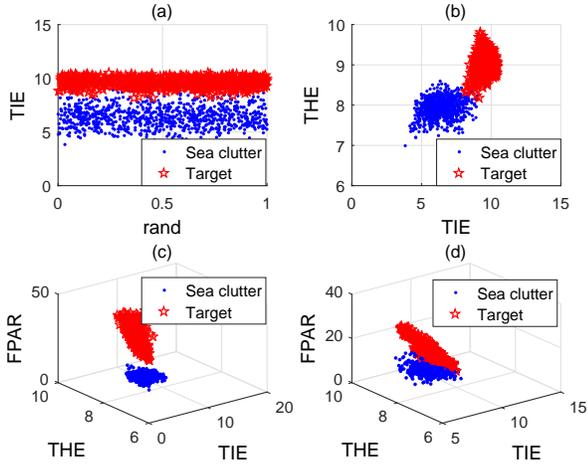}
\centering \caption{Scatter distributions of the target and sea clutter signals. (a) 1-D feature space on $\#54$. (b) 2-D feature space on $\#54$. (c) 3-D feature space on $\#54$. (d) 3-D feature space on $\#30$.}
\label{Fig:Dimension}
\end{figure}

To yield more samples to evaluate the performance of the proposed features, we segment each cell's data of length $2^{17}$ into mutiple small-scale signals of length $D$, given by
\begin{equation}\label{eq:overlapped}
u_j = x(d(j-1)+1:d(j-1)+ D), j = 1,2,\cdots
\end{equation}
where $d$ is a constant to tune the overlapping length among adjacent vectors.
Figs.~\ref{Fig:Describution}(a) and~\ref{Fig:Dimension}(a) exhibit the discriminability of the TIE on $\#54$ under the HH mode through the histogram and scatter distribution, respectively. In both figures, $d$ and $D$ are set to 64 and 4096 (i.e., the observation time is 4096 ms), respectively.
It can be seen that the TIE indeed can be used to distinguish target signals from sea clutter signals, as the TIEs of most target signals are larger than those of sea clutter signals.
However, these two figures also show that effective detection cannot be achieved by only adopting the TIE, as the target and sea clutter signals are highly tangled with each other in some regions.
This is because sea clutter contains spiky pulses in cases of high sea states or low radar grazing angles, which would enlarge the TIEs.

\subsection{Temporal Hurst Exponent}
From \cite{7987714}, the temporal hurst exponent (THE), a widely-used feature to characterize the fractal property of the sea clutter, presents satisfactory discriminability when distinguishing the target from sea clutter.
Inspired by this, we adopt the THE as another feature in our feature space, the calculation procedure of which is described as follows.

Firstly, divide $x$ into $L$ adjacent sub-periods with the same length $\tau = \lfloor\frac{N}{L}\rfloor$ and denote the amplitude set of the $l$-th ($l = 1, 2, \cdots, L$) sub-period by $\{x_{l,1}, x_{l,2}, \cdots, x_{l,\tau}\}$.
Secondly, compute the average amplitude and standard deviation of each sub-period, denoted by $\bar I_l$ and $S_{l}$ for sub-period $l$, respectively.
Let $Y_l =\{Y_{l,1}, Y_{l,2}, \cdots, Y_{l,\tau}\}$ denote the accumulated deviation set of sub-period $l$, where $Y_{l,t}$ is calculated as
$Y_{l,t} = \sum_{k=1}^t(x_{l,k}-\bar I_l), t = 1, 2, \cdots, \tau.$
Define the range of sub-period $l$, denoted by $R_l$, as the difference between the maximum and minimum values of $Y_l$, i.e., $R_{l} = \max(Y_{l,t}) - \min(Y_{l,t})$.

Thirdly, calculate $\frac{R_l}{S_l}$ for all $l\in\{1,2,\cdots,L\}$ for a given $\tau$ and denote the mean value of them by $\frac{R}{S}$.
From \cite{1573749}, $\frac{R}{S}$ shows the fractal feature at a certain time scale range $\tau$, e.g., from 0.1 s to 4 s for the IPIX datasets.
Particularly, $\frac{R}{S}$ is related with the THE, denoted by $H$, by the following equation
\begin{equation}\label{eq:Hurst_four}
\left(\frac{R}{S}\right)_{\tau}=c*{\tau}^H
\end{equation}
where $c$ is a constant independent on $\tau$.
Finally, to expediently calculate $H$, the logarithm operation is taken on both sides of (\ref{eq:Hurst_four}), yielding
\begin{equation}\label{eq:Hurst_five}
\log_2\left(\frac{R}{S}\right)_\tau =  H\log_2\left(\tau\right) + \log_2\left(c\right).
\end{equation}
From (\ref{eq:Hurst_five}), $\log_2\left(\frac{R}{S}\right)_\tau$ is linearly dependent on $\log_2\left(\tau\right)$, and thus $H$ can be readily obtained by the method of first-order least-squares polynomial approximation.

The THEs of the 14 range cells on $\#54$ under the HH mode are plotted in Fig.~\ref{Fig:Describution}(b), from which we can observe that the primary cell has a larger THE than that of the clutter-only cells.
Furthermore, we combine the TIE and THE to construct a 2-D feature space in Fig.~\ref{Fig:Dimension}(b).
Compared with the 1-D feature space (see Fig.~\ref{Fig:Dimension}(a)), the 2-D feature space exhibits better separability.
Nevertheless, there are still some overlaps between the target and sea clutter signals.
As a consequence, it is still necessary to extract additional features for small target detection, which will be described in the next subsection.

\subsection{Frequency Peak to Average Ratio}
To further enhance the discriminability of the feature space, we introduce a frequency-domain feature into it, inspired by the fact that additional spectral information of returned signals that possibly can not be reflected in the time-domain features (e.g., the TIE and HE) can be embedded.
Interestingly, when conducting the Fourier transform on the received signals, we find that the spectrum difference between the target and sea clutter signals exhibits potential discriminability that can be used for detection, as the spectrum of the former mainly distributes over a fluctuant and rough surface while that of the latter more concentrates around a peak.

To quantify this difference, we introduce the frequency peak to average ratio (FPAR) feature, defined as
\begin{equation}\label{eq:PAPR_par}
\text{FPAR}(x)=\frac{{\max\left\{X\left( k \right), {k=1,\cdots,N} \right\}}}{\frac{1}{N}\sum_{k=1}^NX(k)}
\end{equation}
where $X(k)$ is the Fourier transform of the time sequence $x$, given by $X(k)=\sum_{n=1}^{N}x_n\mathrm{e}^{-j\frac{2{\pi}}{N}nk},k = 1,2,\cdots,N.$

Figs.~\ref{Fig:Describution}(c) and~\ref{Fig:Describution}(d) exhibit the FPAR of the target and sea clutter signals, the results in which validate that the simple FPAR does be effective because these two histograms are only slightly overlapped.
Furthermore, we combine the FPAR with the TIE and THE to construct a 3-D feature space, and examine its discriminability through the scatter distribution on $\#54$ in Fig.~\ref{Fig:Dimension}(c).
Compared with the 2-D feature space (see Fig.~\ref{Fig:Dimension}(b)), the 3-D feature space becomes more prominently separable.
However, it is worthwhile to note that some datasets are possibly linearly non-separable in our constructed 3-D feature space, e.g., $\#30$ (see Fig.~\ref{Fig:Dimension}(d)), which indicates that extracting more features does not always result in better separability performance.
Hence, this uncertainty of linear separability should be considered when designing the learning-based detector based on these features, the detailed of which will be described in the next section.

\section{False-Alarm-Rate-Controllable Support Vector Machine Based Detector}
\label{Section:SupportVectorMachine}
Back to the detection problem itself, identifying an object from sea clutter can be naturally regarded as a classification problem.
Based on this fact, this section adopts and elegantly modifies the SVM, a classical and widely-used learning-based binary classifier, to design a detector.
Although SVM-based detectors have been utilized in some existing works to distinguish targets from sea clutter \cite{islam2012artificial,7987714,7775813}, almost all of them directly applied the SVM and did not consider the FAR therein.
However, making the FAR controllable can conveniently regulate detector's sensitivity to the outliers incurred by factors such as the sea spikes and also facilitates to evaluate the performance of different detection algorithms.
It is thus interesting to design a FAR-controllable SVM-based detector when identifying the small targets within sea clutter.

For a sample $i$ in the training dataset, we construct a 3-D feature vector $F_i$ orderly composed by its TIE ($f_{i,1}$), THE ($f_{i,2}$), and FPAR ($f_{i,3}$), i.e., $F_i = [f_{i,1}, f_{i,2}, f_{i,3}]^T$, and use $y_i \in \{+1,-1\}$ to label the class of the target ($+1$) and sea clutter ($-1$).
By this, the $M$ labeled training samples can be represented as $\{(F_i,y_i), i = 1, 2, \cdots, M\}$.
From Figs.~\ref{Fig:Dimension}(c) and~\ref{Fig:Dimension}(d), it is possible that feature vectors of the target and sea clutter are linearly non-separable in the constructed 3-D feature space.
To handle such problem, non-linear kernel functions are introduced into the SVM.
These kernel functions attempt to map $F_i$ into a high-dimensional feature space, where the originally linearly non-separable dataset is shifted to a linearly separable one.
In this letter, we take the radial basis function (RBF) as the kernel function, a prominent choice in SVM-based detectors, defined as follows
\begin{equation}\label{eq:Kernel_One}
k(F_1,F_2) = \mathrm{exp}\left(-\frac{\Vert F_1-F_2\Vert}{2\delta^2}\right).
\end{equation}

After mapping, the next step is to find the hyperplane, i.e., $\omega^TF-b=0$, to separate the target and sea clutter data in the mapped linearly separable high-dimensional feature space according to the max-margin principle.
To determine $\omega$ and $b$, the original SVM, referred to as the $\beta$-SVM in this letter, solves the following quadratic program
\begin{equation}\label{eq:Svm_FirstOptimization}
\begin{aligned}
{\min_{\omega,b,\xi}} ~~& \frac{1}{2}\|\omega\|^2 + \beta\sum_{i=1}^M\xi_i \\
\text{s.t.}         ~~&\text{C1:} ~~ y_i\left[k(\omega, F_i)-b\right] \geq 1-\xi_i, \ i = 1,2,\cdots,M\\
                    ~~&\text{C2:} ~~ \xi_i \geq 0, \ i = 1,2,\cdots,M
\end{aligned}
\end{equation}
where $\xi_i$ is the slack variable and $\beta$ refers to the penalty parameter used to balance the maximization of the margin and the minimization of the error.
Observe that the sea clutter and target signals share the same $\beta$ in the $\beta$-SVM, which implies an assumption that these two classes have the same degrees of toleration to outliers incurred by factors such as the sea spikes.
However, this assumption is possibly not reasonable in practice because the impacts of the outliers on the target and sea clutter signals are usually different.

To deal with this problem, we elegantly modify the $\beta$-SVM to an alternative yet mathematically equivalent version of the $\beta$-SVM, referred to as the FAR-controllable SVM ($P_f$-SVM) in this letter.
Specifically, in the $P_f$-SVM, we introduce two penalty parameters $\beta_0$ and $\beta_1$, respectively for the sea clutter and the target signals, to replace $\beta$ in (\ref{eq:Svm_FirstOptimization}), to control their individual error weights in the quadratic program.
By this, problem (\ref{eq:Svm_FirstOptimization}) is recast to
\begin{equation}\label{eq:Svm_Optimization}
\begin{aligned}
{\min_{\omega,b,\xi}} ~~& \frac{1}{2}\|\omega\|^2 + \sum_{i=1}^M\left(\frac{1-y_i}{2}\beta_0+\frac{1+y_i}{2}\beta_1\right)\xi_i \\
\text{s.t.}         ~~&\text{C1:} ~~ y_i\left[k(\omega, F_i)-b\right] \geq 1-\xi_i, \ i = 1,2,\cdots,M\\
                    ~~&\text{C2:} ~~ \xi_i \geq 0, \ i = 1,2,\cdots,M.
\end{aligned}
\end{equation}

From (\ref{eq:Svm_Optimization}), increasing $\beta_0$ would reduce the FAR for a given $\beta_1$, as by this the obtained hyperplane will tilt toward the target signals and thus less sea clutter signals will be misclassified.
On the other hand, enlarging $\beta_1$ would increase the FAR for a given $\beta_0$, as the hyperplane will be more partial to the sea clutter signals in this case.
Therefore, the modification exploited here not only can enfold the FAR into the SVM-based detector but also facilitates to flexibly control it by simply adjusting $\beta_0$ and $\beta_1$.

In what follows, according to the theory of the SVM, problem (\ref{eq:Svm_Optimization}) can be solved by the sequential minimal optimization (SMO) algorithm in the dual domain \cite{smoo}.
With the obtained hyperplane, i.e., $\omega^TF-b = 0$, the class of an incoming test data $F_j$ can be decided according to the following principle
\begin{equation}\label{eq:Svm_nonlinear}
\left\{
\begin{array}{rcl}
{{ y_j=+1 }}    & & \text{if} \ \ \omega^TF_j-b > 0\\
{{ y_j=-1}}      & &  \text{if} \  \ \omega^TF_j-b \leq 0.\\
\end{array} \right.
\end{equation}

Based on the above discussion, the detailed procedure of our proposed detector is summarized in Algorithm~\ref{Algorithm:ModeSelectionAlgorithm}, in which $\beta_h$ and $\beta_l$ denote the upper and lower bounds of $\beta_0$, respectively.
The algorithm runs in two stages.
In the first stage (Lines 3--5), obtain the hyperplane with the given parameters. Then, use this hyperplane to classify the training data and calculate the actual FAR $P_F$.
In the second stage (Lines 6--17), adopt the bi-section method to adjust $\beta_0 $ by comparing $P_F$ with the user-defined FAR $P_f$.
These two stages will be executed iteratively until the difference between $P_F$ and $P_f$ is lower than the predefined threshold $\eta$.

\begin{algorithm}[!t]
\caption{FAR-Controllable SVM-Based Detector.}
\begin{algorithmic}[1] \label{Algorithm:ModeSelectionAlgorithm}
\STATE \textbf{Initialization}\\
\begin{itemize}
\item Set the FAR $P_f$ and the threshold $\eta$ (e.g., 0.0001).
\item Set $\beta_h = 2, \beta_l=0$, $\beta_0 = 1$, $\beta_1 = 1$, and $P_F = 1$.
\end{itemize}
\WHILE {$|P_F - P_f|$ > $\eta$}
\STATE Solve (\ref{eq:Svm_Optimization}) to obtain $\omega$ and $b$.
\STATE Determine the class of the training data by (\ref{eq:Svm_nonlinear}).
\STATE Calculate the FAR, defined as \\
$P_F\! = \!\frac{\text{The number of misclassified sea clutter samples}}{\text{The total number of sea clutter samples in training dataset}}\!\times100\%\!$.
\IF {$P_F$   = $P_f$}
\STATE Break.
\ELSE
 \IF{$P_F$ < $P_f$}
 \STATE Set $\beta_h$ = $\beta_0$ and $\beta_0$ = $\frac{\beta_h+\beta_l}{2}$.
 \ENDIF
\ELSE
 \IF{$P_F$ > $P_f$}
 \STATE Set $\beta_l$ = $\beta_0$ and $\beta_0$ = $\frac{\beta_h+\beta_l}{2}$.
 \ENDIF
\ENDIF
\ENDWHILE
\STATE Calculate the detection probability defined as \\
$P_d = \frac{\text{The number of correctly-classified target samples}}{\text{The total number of target samples in testing dataset}}\times100\%$.
\end{algorithmic}
\end{algorithm}

\section{Experimental Results} \label{Section:Simulation}
In this section, we use the 10 datasets mentioned in Section~\ref{Section:SystemModel} to evaluate the performance of our proposed detector.

Consider that sufficient signal samples are needed to train the learning-based detector, the overlapped segmentation is thus adopted under the partition rule presented in (\ref{eq:overlapped}), with the parameters set to $d = 64$ and $D = 4096$, respectively.
By this, we could yield 1984 target samples and more than 20000 sea clutter samples for each dataset.
Then, we divide the obtained samples into two groups, one for training composed of a half of the target samples and all the clutter-only samples and the other for testing composed of the rest of target samples.

\begin{figure}[t]
\centering \leavevmode \epsfxsize=3.5 in
\epsfbox{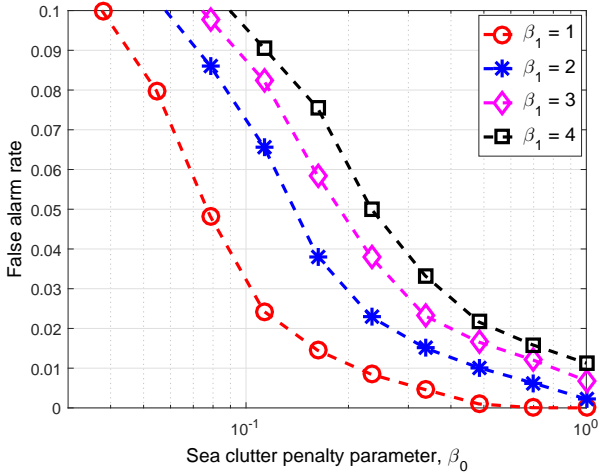}
\centering \caption{FAR versus $\beta_0$ on $\#17$ under the VV mode.}
\label{Fig:Result_SCR}
\end{figure}

To verify whether our proposed detector can flexibly tune the FAR or not, we test its performance on $\#17$ under the VV mode.
Fig.~\ref{Fig:Result_SCR} illustrates how the two introduced penalty parameters $\beta_0$ and $\beta_1$ impact the FAR.
From the figure, it is obtained that a higher $\beta_0$ corresponds to a lower FAR for a given target penalty parameter $\beta_1$ and a higher $\beta_1$ results in a larger FAR for a given $\beta_0$.
Hence, our proposed detector can flexibly control the FAR by simply adjusting $\beta_0$ and $\beta_1$.

Furthermore, to evaluate the performance of our proposed detector under varying detection environments, we compare it with two classical detectors, the tri-feature detector \cite{6850164} and the fractal-based detector \cite{1573749}.
Firstly, we compare their detection performance under different SCR situations in Table~\ref{Table:3Dcomparison2}.
It can be observed that our proposed detector can attain better detection performance than the other two in both the high and low SCR cases.
For example, our proposed detector improves the detection probability by $58\%$ and $19\%$ compared with the fractal-based and tri-feature detectors, respectively, in the case of SCR $=$ -2 dB.

\begin{figure}[t]
\centering \leavevmode \epsfxsize=3.5 in
\epsfbox{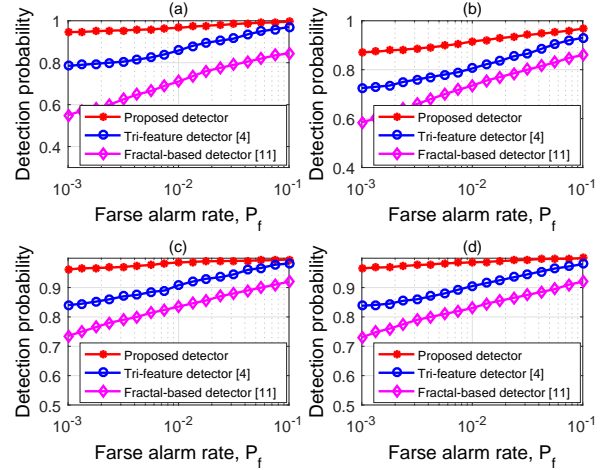}
\centering \caption{Comparisons of the detection probability under (a) HH, (b) VV, (c) HV, and (d) VH modes, where the observation time is set to 4096 ms.}
\label{Fig:svm_result}
\end{figure}

Secondly, we compare their detection performance at different FARs in Fig.~\ref{Fig:svm_result}, where the detection probability is obtained by first calculating the detection probabilities of all the datasets and then taking an average on them.
It can be seen that, although the detection probabilities of these three detectors all increase with the FAR, our proposed detector always achieves better detection performance than the other two either in high or low FAR cases.
For instance, our proposed detector improves the detection probability by $16\%$ and $40\%$ compared with the tri-feature detector and fractal-based detector under the HH mode, respectively, when the FAR is 0.001.
\begin{table}[!t]
\vspace{-0.2cm}  
\setlength{\abovecaptionskip}{-0.1cm}   
\setlength{\belowcaptionskip}{-1cm}   
\centering
\caption{Comparisons of the detection probability $(P_f=0.001)$.}
\label{Table:3Dcomparison2}
\begin{tabular}{ccccc}
\toprule[1pt]
 \multicolumn{1}{c}{\multirow{2}{*}{\textbf{Methods}}} & \multicolumn{2}{c}{\textbf{$P_d$ (HH mode)}}  \\ \cline{2-3}
 \multicolumn{1}{c}{} & \multicolumn{1}{c}{\textbf{SCR=-2 dB}} & \multicolumn{1}{c}{\textbf{SCR=17 dB}}   \\
 \multicolumn{1}{c}{\textbf{Proposed detector}} & \multicolumn{1}{c}{\textbf{76}} & \multicolumn{1}{c}{\textbf{99}}  \\
 \multicolumn{1}{c}{\textbf{Tri-feature detector~\cite{6850164}}} & \multicolumn{1}{c}{57} & \multicolumn{1}{c}{99}   \\
 \multicolumn{1}{c}{\textbf{Fractal-based detector \cite{1573749}}} & \multicolumn{1}{c}{18} & \multicolumn{1}{c}{79}   \\
\bottomrule[1pt]
\end{tabular}
\end{table}

\section{Conclusions} \label{Section:Conclusions}
Taking the varying detection environments into account, this letter has investigated the problem of detecting small targets floating on sea surface.
For this, we have first extracted three discriminative features and then designed a SVM-based detector that can flexibly tune the FAR.
Experimental results have verified the superiority of our proposed detector over several existing detectors in both low SCR and low FAR cases.
\bibliographystyle{IEEEtran}
\bibliography{IEEEabrv,LatexWritingModel_Reference}

\end{document}